\documentclass[%
 reprint,
 superscriptaddress,
 amsmath,amssymb,
 aps,
 prl,
]{revtex4-2}

\usepackage{physics}
\usepackage{graphicx}
\usepackage{dcolumn}
\usepackage{bm}
\usepackage{booktabs} 
\usepackage{float}
\usepackage{placeins}
\usepackage{fancyhdr}

\usepackage{xcolor}
\newcommand{\TUIlmenau}{Institute of Physics and Institute of Micro- and Nanotechnologies, Technische Universit\"at Ilmenau, 98693 Ilmenau, Germany}

\begin{document}

\title{Quantum nuclear and band-dispersion effects recover near-UV absorption in short-hydrogen-bonded organic crystals
}

\author{Jonas H{\"a}nseroth}
\affiliation{\TUIlmenau}

\author{Max Gro{\ss}mann}
\affiliation{\TUIlmenau}

\author{Malte Grunert}
\affiliation{\TUIlmenau}

\author{Ali Hassanali}
\affiliation{Condensed Matter and Statistical Physics, The Abdus Salam International Centre for Theoretical Physics (ICTP), 34151 Trieste, Italy}

\author{Erich Runge}
\affiliation{\TUIlmenau}

\author{Christian Dre{\ss}ler}
\affiliation{\TUIlmenau}

\author{Muhammad Nawaz Qaisrani}
\email{muhammad-nawaz.qaisrani@tu-ilmenau.de}
\affiliation{\TUIlmenau}

\date{\today}

\begin{abstract}
Near-UV optical absorption is increasingly reported in hydrogen-bonded organic and biomolecular materials lacking aromatic or extended $\pi$-conjugated chromophores, yet its microscopic origin remains unresolved and electronic-structure calculations often overestimate experimental absorption onsets. Here, we combine machine-learned interatomic potentials for large-scale classical and quantum nuclear sampling with periodic excited-state calculations to address this discrepancy in \textit{L}-pyroglutamine ammonium, an experimentally established glutamine-derived crystal containing a well-resolved short hydrogen bond and exhibiting non-aromatic near-UV optical response. Using controlled \textit{in silico} ion substitutions that vary the surrounding hydrogen-bond environment while preserving this scaffold, we compute optical spectra from configurations sampled along classical and quantum nuclear trajectories using hybrid-functional time-dependent density functional theory. We show that nuclear quantum effects stabilise proton-sharing configurations that are strongly suppressed classically, redshifting the lowest bright excitations by $0.5$--$0.8~\mathrm{eV}$ and raising the fraction of configurations with bright excitations below $6~\mathrm{eV}$ from approximately $3\%$ to approximately $30\%$. Explicit Brillouin-zone sampling provides a further, mechanistically distinct redshift of $0.5$--$1.1~\mathrm{eV}$, reflecting modest but significant indirect electronic character. Only when both effects are incorporated does the calculated onset recover the experimental $3.8$--$4.5~\mathrm{eV}$ range. These results establish quantum proton fluctuations and reciprocal-space convergence as cooperative but physically distinct ingredients required for predictive optical spectroscopy of strongly hydrogen-bonded molecular materials.
\end{abstract}

\maketitle

\section{Introduction}

Near-UV optical excitations in organic and biomolecular materials are conventionally associated with aromatic residues or extended $\pi$-conjugated chromophores supporting delocalized excited states. \cite{teale1957ultraviolet,lakowicz2006principles}
However, over the past decade, measurable near-ultraviolet (UV) absorption and visible emission have been reported in a broad range of systems lacking classical conjugation, including peptide and amyloid assemblies, amino-acid crystals, molecular aggregates, and non-conjugated polymers. \cite{pinotsi2013label,pinotsi2016proton,grisanti2020toward,morzan2022non,balasco2023comprehensive,niyangoda2017carbonyl,arnon2021off,stephens2021short,ye2017non,tang2021nonconventional,liu2023clusteroluminescence}
Across these systems, low-energy optical absorption often emerges or intensifies upon supramolecular organization through crystallization, aggregation, or extended hydrogen-bond networks, consistent with broader concepts such as clusteroluminescence and clusterization-triggered emission in nonconventional luminophores. \cite{ye2017non,zhou2017oligo,tang2021nonconventional,zhang2021through}
These observations suggest that the relevant excited states are collective condensed-phase phenomena shaped by intermolecular electronic interactions, electrostatics, hydrogen-bond cooperativity, and nuclear motion. \cite{morzan2022non,stagi2021thermal}
Despite extensive experimental observations, the microscopic origin of these excitations remains unresolved, while existing theoretical calculations frequently overestimate experimental absorption onsets.

Several mechanisms have been proposed to explain these excitations, including red-shifted carbonyl-localized transitions, intermolecular charge-transfer states stabilized by hydrogen-bond networks, and collective electrostatic effects in densely packed molecular assemblies. \cite{pinotsi2016proton,morzan2022non,stagi2022root}
In peptide and amyloid systems, low-energy optical absorption correlates strongly with extended hydrogen-bond networks characteristic of cross-$\beta$ structures \cite{grisanti2017computational,balasco2023comprehensive,wang2025intrinsic}, while first-principles simulations suggest that proton motion along hydrogen bonds can modulate optically active electronic transitions even in the absence of aromatic chromophores.\cite{pinotsi2016proton,morzan2022non}
Yet statistically converged optical calculations that simultaneously incorporate thermal disorder, nuclear quantum effects (NQEs), and Brillouin-zone (BZ) convergence remain computationally prohibitive for large periodic molecular systems. 
As a result, most previous theoretical studies relied on geometry-optimized structures, limited configurational sampling, or $\Gamma$-point approximations,\cite{pinotsi2016proton,grisanti2017computational,miron2023carbonyl,stephens2021short,jong2019low}
and the quantitative consequences of these approximations for low-energy optical absorption remain largely unexplored. 
Moreover, the structural heterogeneity of biomolecular aggregates obscures identification of minimal emitting motifs and complicates quantitative structure--property relationships \cite{balasco2023comprehensive}, motivating the use of structurally well-defined crystalline systems to isolate the fundamental photophysical mechanisms governing non-aromatic optical response.

A prototypical system that addresses these challenges is the glutamine-derived molecular crystal L-pyroglutamine ammonium (L-pyro-amm), formed through mild thermal conversion of L-glutamine. \cite{stephens2021short} Experimentally, L-pyro-amm exhibits near-UV absorption and visible fluorescence despite the absence of aromatic chromophores. 
Single-crystal diffraction revealed a short hydrogen bond (SHB), defined here by an O--O distance of approximately $2.45$~{\AA} and partial proton-sharing character, located adjacent to an ammonium ion and forming a minimal, structurally well-resolved hydrogen-bond motif. 
Subsequent theoretical studies suggested that this strong hydrogen bonding suppresses non-radiative decay pathways by restricting carbonyl elongation and limiting access to conical intersections, a mechanism termed the "carbonyl lock". \cite{miron2023carbonyl}
These studies established the importance of hydrogen-bond geometry in stabilizing excited states, but primarily focused on excited-state relaxation pathways rather than on the microscopic origin of the low-energy optical absorption onset itself.

More recently, we showed using ab-initio path-integral molecular dynamics (PIMD) simulations that NQEs fundamentally reorganize the SHB ground-state ensemble in L-pyro-amm by symmetrizing proton transfer and enhancing electronic polarization within the hydrogen-bond network \cite{qaisrani2025acid}.
These findings raise a central unresolved question: whether quantum proton fluctuations merely modify structural stability or fundamentally reshape the low-energy optical excitation landscape associated with non-aromatic optical absorption.

Previous multiscale spectroscopy studies have shown that NQEs can substantially influence excitation energies, band-gap renormalization, spectral broadening, and optical line shapes in hydrogen-bonded systems.\cite{feher2021multiscale,law2018importance,tsuru2025nuclear,chen2016ab,berrens2024nuclear,pinotsi2016proton,bischoff2021band} However, their quantitative impact on the low-energy optical absorption of periodic hydrogen-bonded molecular crystals remains poorly understood. Moreover, even when nuclear sampling is treated accurately, optical spectra are commonly evaluated using only the $\Gamma$ point,\cite{pinotsi2016proton,grisanti2017computational,miron2023carbonyl,stephens2021short,jong2019low} implicitly assuming that reciprocal-space effects are negligible. The combined influence of thermal disorder, quantum nuclear fluctuations, and Brillouin-zone sampling on optical absorption onsets in hydrogen-bonded molecular materials therefore remains largely unknown. Addressing this problem requires configurational ensembles far larger than those accessible to direct first-principles simulations. Recent advances in machine-learned interatomic potentials (MLIPs) trained on \textit{ab initio} data provide a practical route to overcome this limitation while retaining near-first-principles accuracy.\cite{behler2007generalized,batatia2022mace}

Here, we combine MLIP-enabled large-scale classical and PIMD nuclear sampling with periodic time-dependent density functional theory (TDDFT)\cite{runge1984density} calculations to investigate low-energy optical excitations in the short-hydrogen-bonded molecular crystal L-pyro-amm. MLIPs are used to generate extensive classical and quantum nuclear ensembles, while optical spectra are computed separately from configurations extracted from these ensembles using hybrid-functional excited-state calculations. Starting from geometry-optimized crystal structures, we systematically incorporate thermal disorder and NQEs into configurational ensembles, while BZ convergence is assessed through explicit $k$-point sampling calculations on representative configurations. To probe the influence of local hydrogen-bond cooperativity, we further perform controlled \textit{in silico} ion substitutions by replacing the original NH$_4^+$ ion with chemically distinct K$^+$ and H$_3$O$^+$ ions while preserving the intrinsic SHB scaffold. This framework enables a systematic assessment of how thermal disorder, quantum nuclear fluctuations, and reciprocal-space sampling influence the low-energy optical absorption of short-hydrogen-bonded molecular crystals. By isolating and quantifying these contributions, we establish a physically grounded basis for understanding the origin of non-aromatic near-UV optical response and for assessing the level of theory required for predictive optical spectroscopy in hydrogen-bonded molecular materials.

\begin{figure*}[t]
    \centering
    \includegraphics[clip=true, width=\textwidth]{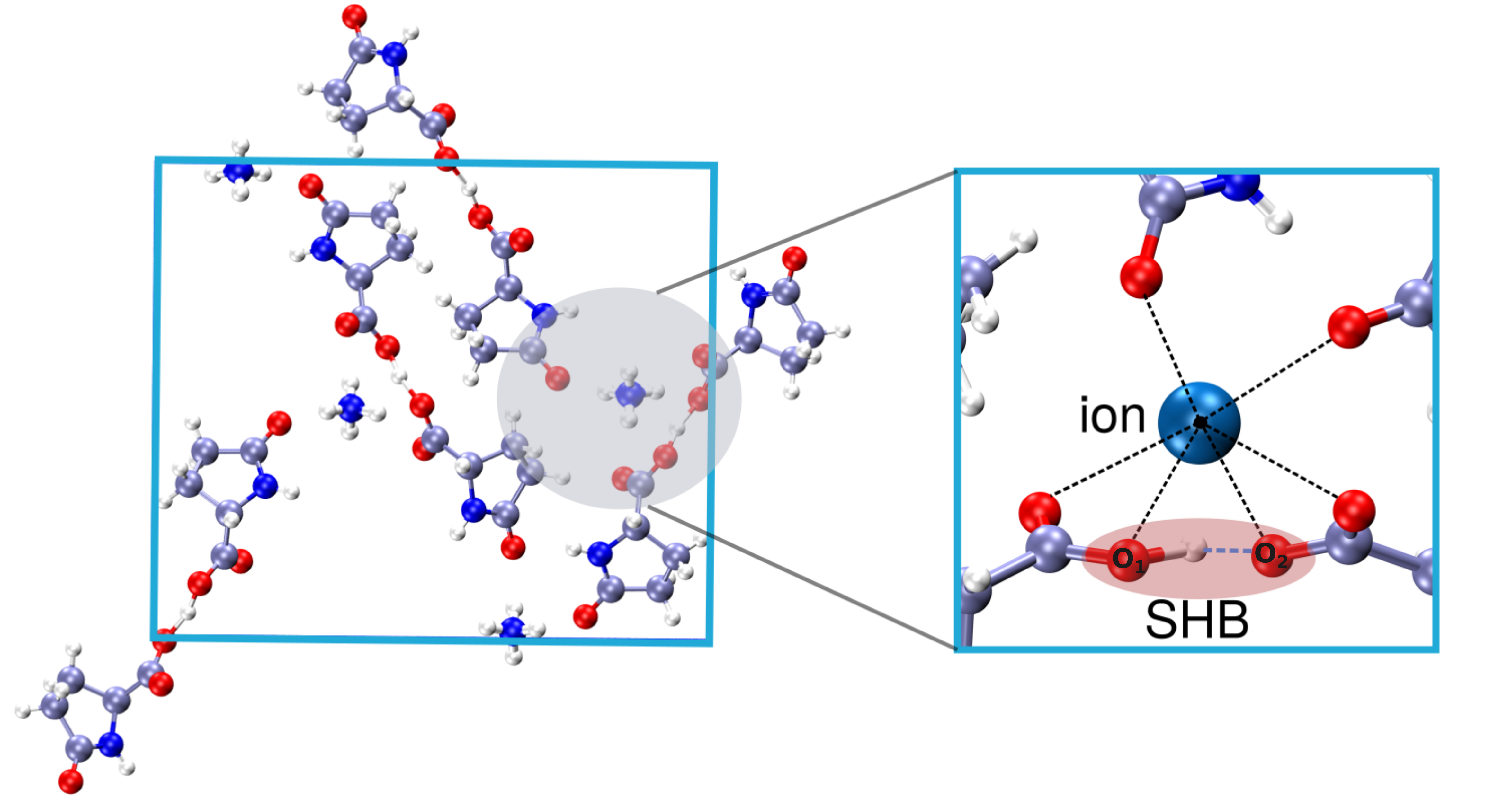}
    \caption{\textbf{Preserved short-hydrogen-bond scaffold and ion-dependent local hydrogen-bond environments in glutamine-derived crystals.}
    Left: experimentally resolved crystal structure of L-pyro-amm containing four SHB motifs in the unit cell connecting the pyroglutamate and pyroglutamic acid groups.
    Right: zoomed view of the local ionic environment surrounding the SHB scaffold, with the cation shown occupying the local coordination cavity adjacent to the SHB network.}
    \label{structure}
\end{figure*}

\section{Results}

\subsection{Crystal structure and \textit{in silico} ion substitution}

A central question in this work is whether the low-energy optical response is governed primarily by the intrinsic SHB motif or by the surrounding ionic and hydrogen-bond environment. To disentangle these contributions, we constructed a series of \textit{in silico} ion-substituted crystals in which the original NH$_4^+$ ion of L-pyro-amm was replaced by K$^+$ or H$_3$O$^+$ while preserving the underlying SHB scaffold.

Figure~\ref{structure} shows the experimentally resolved L-pyro-amm crystal structure together with the local ionic environment surrounding the SHB motif. In the experimental crystal, the NH$_4^+$ ion occupies a coordination cavity adjacent to the SHB network and forms multiple hydrogen-bond interactions with nearby oxygen atoms.\cite{qaisrani2025acid} Replacing NH$_4^+$ with K$^+$ or H$_3$O$^+$ therefore provides a controlled means of modifying the local electrostatic and hydrogen-bond environment surrounding the SHB without directly altering the molecular framework responsible for proton transfer.

Despite the substantial chemical differences between the three ions, unconstrained variable-cell DFT relaxations preserve both the orthorhombic crystal symmetry and the SHB motif connecting the pyroglutamate and pyroglutamic acid units. Optimized lattice volumes differ by only $\sim2\%$ across the three compositions (see Supplementary Note~1), indicating that the substitutions induce only minor structural perturbations. The primary response occurs locally around the ionic site through reorganization of hydrogen-bond coordination.

The three ions generate qualitatively distinct local environments around the SHB motif. K$^+$ acts primarily as an electrostatic stabilizer and does not introduce additional directional hydrogen bonds. NH$_4^+$ forms multiple N--H--O interactions with neighboring oxygen atoms, whereas H$_3$O$^+$ introduces additional O--H--O hydrogen-bond and proton-sharing interactions. Together, K$^+$ and H$_3$O$^+$ therefore span two limiting cases of the local environment: a predominantly electrostatic stabilizer and a strongly hydrogen-bonding proton donor.

The resulting ion series provides a controlled platform for testing whether proton-sharing behavior and low-energy optical response are intrinsic consequences of the SHB itself or depend sensitively on the surrounding ionic environment. In the following section, we examine how thermal fluctuations and NQEs influence proton localization and proton sharing across these three distinct local environments.

\subsection{NQEs drive a common proton-sharing ensemble across distinct ionic environments}

\begin{figure}[ht]
    \centering
    \includegraphics{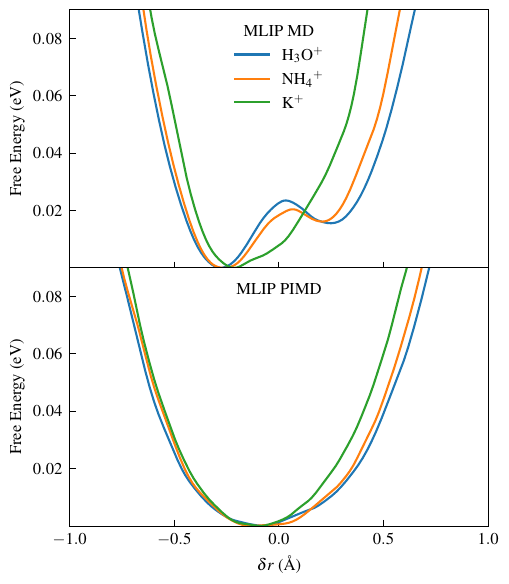}
    \caption{\textbf{NQEs suppress ion-dependent proton localization and stabilize proton-sharing configurations.}
    Proton-transfer free-energy profiles $F(\delta r)$ for NH$_4^+$, K$^+$, and H$_3$O$^+$ crystals obtained from classical MLIP molecular dynamics (top) and MLIP PIMD simulations including NQEs (bottom) at 300~K. Classical ensembles exhibit pronounced ion-dependent asymmetry, whereas NQEs drive all three systems toward more symmetric proton-sharing free-energy profiles centered near $\delta r=0$.}
    \label{free_energy}
\end{figure}

The preserved SHB scaffold allows us to directly assess whether proton localization remains sensitive to the local ionic environment once NQEs are included. Figure~\ref{free_energy} compares free-energy profiles along the proton-transfer coordinate $\delta r$ (see Eqs.~1 and 2 in the Methods) obtained from MLIP MD and MLIP PIMD simulations for the three ion-substituted crystals.

Under classical sampling, the proton-transfer landscape depends strongly on the local ionic environment. The NH$_4^+$ and H$_3$O$^+$ crystals exhibit relatively shallow asymmetric double-well profiles characteristic of low-barrier proton transfer, whereas the K$^+$ crystal displays a more asymmetric profile in which the proton preferentially localizes toward one oxygen atom. Notably, the NH$_4^+$ crystal exhibits the same shallow double-well topology previously identified for the experimental L-pyro-amm crystal, indicating that the MLIP-generated ensembles reproduce the essential SHB behavior established in earlier first-principles simulations.\cite{qaisrani2025acid} Importantly, all three classical profiles exhibit only shallow barriers of a few tens of meV, making them highly susceptible to quantum delocalization effects. Thus, under classical sampling, the surrounding ion strongly influences proton localization through modifications of the local hydrogen-bond environment.

In contrast, NQEs drive all three systems toward more symmetric free-energy profiles centered near $\delta r = 0$, indicating enhanced stabilization of proton-sharing configurations. At the same time, the pronounced ion-dependent asymmetry observed in the classical ensembles is substantially reduced. Proton sharing therefore emerges as a robust feature of the SHB under NQEs, despite the chemically distinct local hydrogen-bond environments surrounding the proton-transfer motif.

To quantify this reduction in ion dependence, we analyzed ensemble-averaged proton-transfer descriptors obtained from the long-timescale MD and PIMD trajectories. Under classical sampling, the proton-localization magnitude $\langle |\delta r| \rangle$ varies across the three ion substitutions, whereas the quantum ensembles exhibit markedly reduced variability. The ion-dependent spread in $\langle |\delta r| \rangle$ decreases by approximately $60\%$ upon inclusion of NQEs. A similar reduction is observed for the proton-transfer fluctuation width $\sigma(\delta r)$, whose ion-dependent spread decreases from $0.037$~{\AA} in the classical ensembles to $0.016$~{\AA} in the quantum ensembles. The donor--acceptor oxygen distance exhibits a smaller but systematic reduction in ion-dependent variability. Together, these results demonstrate that NQEs suppress counter-ion-specific proton localization and stabilize a more homogeneous proton-sharing ensemble within the SHB. Detailed statistical analysis, convergence tests, and uncertainty estimates obtained from block averaging are provided in Supplementary Note~6.

\begin{figure*}[ht]
    \centering
    \includegraphics[width=\textwidth]{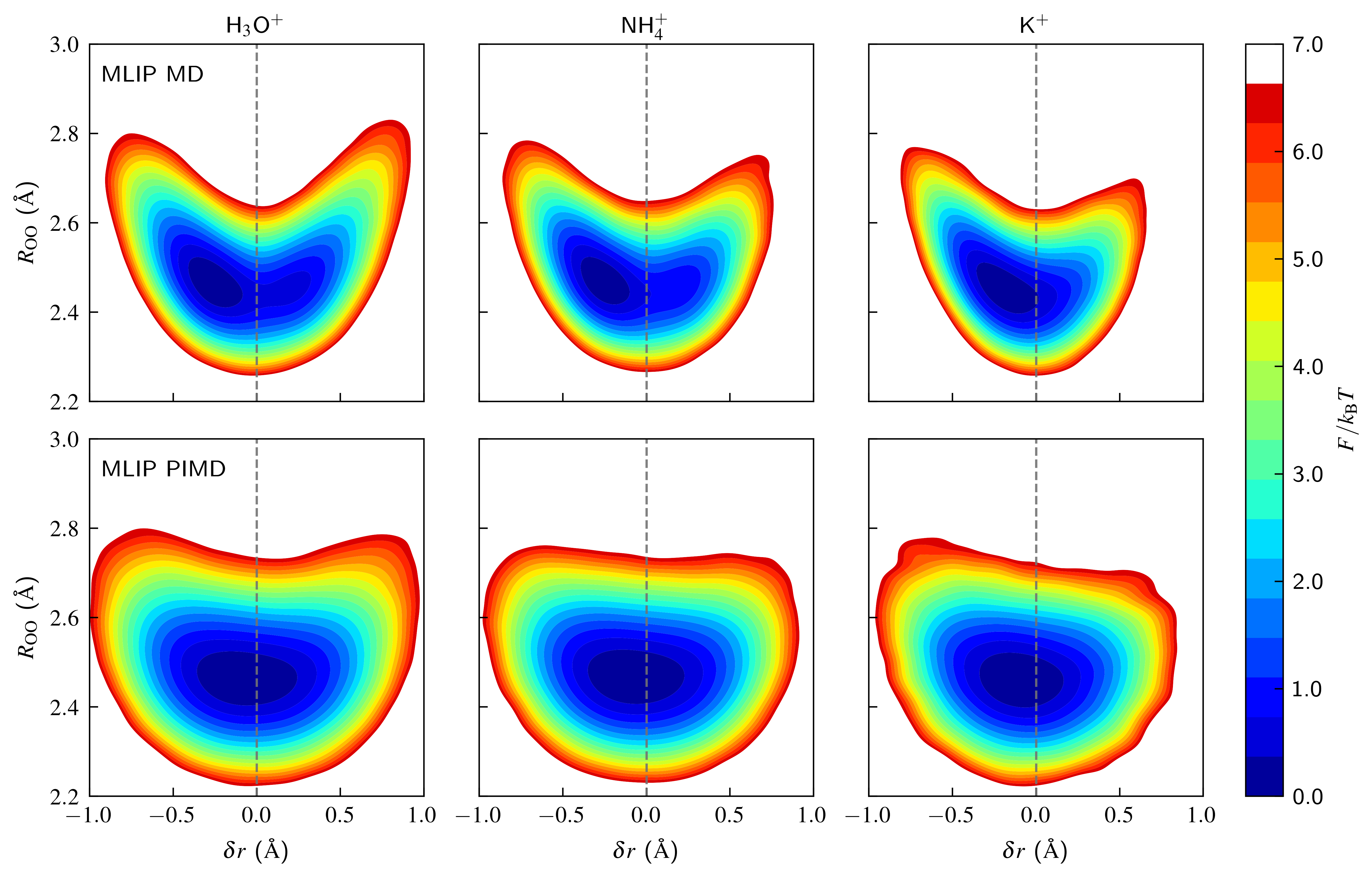}
    \caption{\textbf{NQEs reduce coupling between proton transfer and donor--acceptor fluctuations.}
    Two-dimensional free-energy surfaces $F(\delta r,R)$ describing coupled proton-transfer and donor--acceptor fluctuations within the SHB. Top row: classical MLIP MD simulations. Bottom row: MLIP PIMD simulations including NQEs. The color scale is given in units of $F/k_{\mathrm{B}}T$ at $T=300$~K. Vertical dashed lines indicate the symmetric proton-sharing position $\delta r = 0$. Classical MLIP MD exhibits ion-dependent asymmetry and pronounced coupling between proton displacement and donor--acceptor compression, whereas MLIP PIMD produces more symmetric proton-sharing basins with reduced ion dependence and weaker $\delta r$--$R$ correlation.
}

    \label{fig:2d_ptc_r}
\end{figure*}

The two-dimensional free-energy surfaces $F(\delta r,R)$ further reveal how proton motion couples to donor--acceptor rearrangement (Fig.~\ref{fig:2d_ptc_r}). Classical ensembles exhibit elongated and strongly ion-dependent basins, indicating pronounced coupling between proton displacement and O--O compression. In contrast, PIMD produces more compact and nearly symmetric basins with significantly reduced diagonal correlation between $\delta r$ and $R$. Together, these results show that although the surrounding ions strongly influence proton localization under classical sampling, NQEs drive all systems toward a common proton-sharing SHB ensemble that is only weakly dependent on ion identity. This raises the possibility that low-energy optical excitations may likewise be governed primarily by the quantum proton-sharing character of the SHB rather than by details of the local ionic environment.

\subsection{NQEs stabilize low-energy optically bright configurations}

\begin{figure*}[ht]
    \centering
    \includegraphics{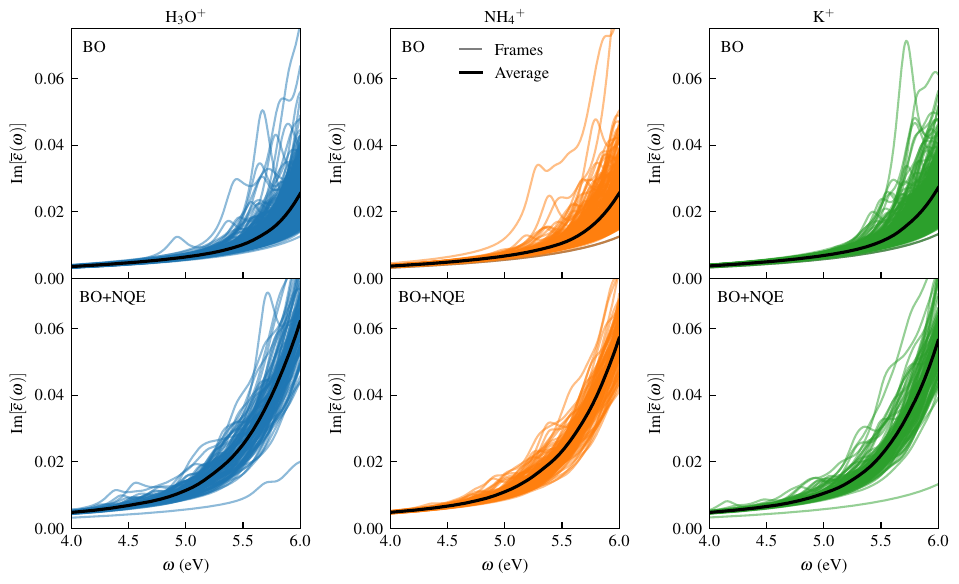}
    \caption{\textbf{Quantum nuclear ensembles stabilize low-energy optically bright configurations.}
    Ensemble-resolved TDDFT absorption spectra for ion-substituted crystals comparing configurations sampled from classical ensembles (top row) and quantum ensembles including NQEs (bottom row). Columns correspond to H$_3$O$^+$, NH$_4^+$, and K$^+$. Thin curves show spectra of individual configurations, whereas thick black curves denote ensemble averages. NQEs substantially increase the population of low-energy optically bright configurations and redistribute spectral weight toward lower excitation energies.
}

    \label{tddft_ensembles}
\end{figure*}

The emergence of a common proton-sharing SHB ensemble under NQEs raises a natural question: how does this quantum structural reorganization influence the low-energy optical excitation landscape?
TDDFT spectra were computed for configurations sampled from the classical and quantum nuclear trajectories for each ion-substituted crystal. 
Figure~\ref{tddft_ensembles} shows the resulting ensemble-resolved dielectric response, where thin curves denote spectra of individual configurations and thick black curves denote ensemble averages.
The optical spectra exhibit substantial configurational heterogeneity, indicating that the low-energy absorption onset is controlled by a relatively small subset of low-energy bright configurations whose population depends sensitively on the underlying nuclear ensemble. 

Under classical sampling, appreciable optical intensity occurs primarily above approximately $5$--$6$~eV, while bright low-energy excitations remain strongly suppressed. 
Inclusion of NQEs qualitatively reorganizes the excitation landscape. 
Across all ion substitutions, the quantum ensembles exhibit persistent low-energy optical intensity in the 4--5 eV region arising from an enhanced population of proton-sharing configurations identified in the preceding section.

Using the ensemble optical descriptors defined in the Methods, we quantify the quantum-induced spectral reorganization in Tab.~\ref{tab:summary_optical_reorganization}.
 
Across all systems, inclusion of NQEs redshifts the lowest bright excitation energies by approximately $0.5$--$0.8$~eV relative to the corresponding classical ensembles.
Most notably, the fraction of sampled configurations exhibiting bright excitations below $6$~eV increases from only $\sim3\%$ in the classical ensembles to approximately $\sim30\%$ in the quantum ensembles. 
Thus, NQEs do not merely broaden the spectra, but instead strongly redistribute the statistical population of optically bright configurations toward lower excitation energies.

The magnitude of this optical enhancement systematically follows the degree of local hydrogen-bond cooperativity surrounding the SHB scaffold. The strongest quantum-induced spectral reorganization occurs for the H$_3$O$^+$ crystal, whereas the weakest enhancement occurs for K$^+$, which primarily provides electrostatic stabilization with minimal directional hydrogen bonding. Importantly, however, all three ion substitutions exhibit the same qualitative quantum-induced redistribution of low-energy optical intensity despite their chemically distinct local environments. This persistence indicates that the SHB scaffold itself governs the low-energy excitation landscape, while the surrounding ions primarily modulate the degree of proton sharing and hydrogen-bond cooperativity.

To assess the robustness of this conclusion with respect to the electronic-structure treatment, we performed additional calculations within the independent-particle approximation and selected B3LYP benchmark calculations. Both approaches reproduce the same qualitative quantum-induced enhancement of low-energy optical intensity and the corresponding redistribution of optically bright configurations (Supplementary Notes 4 and 5), indicating that the observed trends are not specific to the chosen excited-state methodology.

\begin{figure}[ht]
\centering
\includegraphics{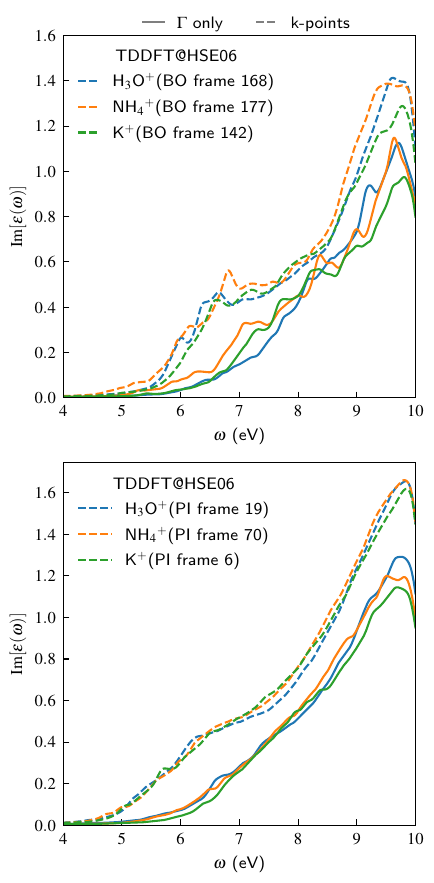}
\caption{\textbf{Explicit BZ sampling systematically lowers the optical absorption onset.}
Comparison of $\Gamma$-point and explicitly BZ-sampled TDDFT absorption spectra for configurations exhibiting the lowest-energy absorption onset within the sampled ensembles. Explicit BZ sampling systematically lowers the absorption onset and enhances low-energy spectral weight for all ion substitutions and for both classical (top) and quantum (bottom) configurations. Similar BZ-induced redshifts are observed for configurations representative of the ensemble-average optical spectra (Supplementary Note~5).
}

\label{fig:bz_comparison}
\end{figure}

\subsection{Brillouin-zone sampling further redshifts the absorption onset}

The ensemble optical calculations discussed above were performed at the $\Gamma$ point because statistically converged hybrid-functional TDDFT calculations with explicit BZ sampling remain computationally prohibitive for the present 144-atom molecular crystal, as each calculation is approximately 200--300 times more expensive than the corresponding $\Gamma$-point calculation.
Despite the large unit cell, analysis of the electronic structure reveals residual band dispersion and a weakly indirect electronic gap away from the $\Gamma$ point (Supplementary Note~5). These features indicate that the low-energy optical excitations remain sensitive to BZ sampling and suggest that conventional $\Gamma$-point calculations may systematically overestimate the optical absorption onset even in large molecular crystals.

To assess the magnitude of this effect, selected configurations from both classical and quantum ensembles were recomputed using explicit BZ sampling. Two classes of structures were analyzed for each ion substitution: configurations representative of the ensemble-average optical spectra and configurations exhibiting the lowest optical absorption onset within the sampled ensembles. Figure~\ref{fig:bz_comparison} compares the corresponding $\Gamma$-point and explicitly BZ-sampled spectra for the low-onset configurations, while analogous comparisons for ensemble-average structures are provided in Supplementary Note~5. The full set of benchmark calculations is summarized in Supplementary Table~S4.

\begin{figure}[ht]
\centering
\includegraphics{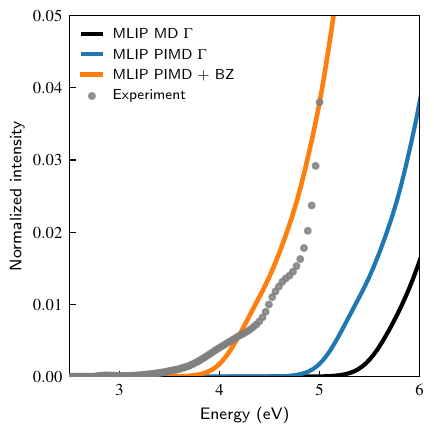}
   \caption{\textbf{Quantum nuclear sampling and explicit BZ sampling jointly recover the experimental near-UV absorption onset.}
   Comparison between the experimentally measured absorption spectrum and ensemble-averaged TDDFT optical spectra for the L-pyro-NH$_4^+$ crystal. Ensemble-averaged spectra obtained from classical MLIP MD (black) and quantum MLIP PIMD (blue) configurations were computed at the $\Gamma$ point from the imaginary part of the dielectric function, Im[$\varepsilon(\omega)$]. The orange curve shows the MLIP PIMD spectrum after applying the mean BZ-induced redshift (0.96~eV) obtained from explicit $k$-point TDDFT benchmark calculations on twelve representative configurations. Experimental data from Ref.~\cite{stephens2021short} are shown as gray markers. The experimental near-UV absorption onset lies approximately in the $3.8$--$4.5$~eV range. Experimental and theoretical spectra were baseline-corrected and independently normalized to facilitate comparison of the absorption-onset region. 
    No fitting to experiment was performed.}
    \label{fig:exp_overlay}
\end{figure}

Explicit BZ sampling systematically lowers the optical absorption onset for all ion substitutions and all sampled structures examined, including both classical and quantum ensembles as well as both ensemble-average and low-onset configurations (Fig.~\ref{fig:bz_comparison}; Supplementary Note~5). The resulting onset redshifts range from approximately $0.5$ to $1.1$~eV and are consistently accompanied by redistribution of spectral weight toward lower excitation energies. While this full benchmark set spans both classical and quantum configurations, the mean correction applied to the quantum ensemble in Fig.~\ref{fig:exp_overlay} ($0.96$~eV) is obtained from the quantum-configuration subset alone, reported separately as $\Delta E_k$ in Tab.~\ref{tab:summary_optical_reorganization}.

Importantly, the magnitude of the BZ correction remains relatively consistent across ion substitutions, ensemble types, and sampled structures. Despite thermal fluctuations and differing local hydrogen-bond environments, the onset shifts remain confined to a comparatively narrow range, suggesting that the weakly indirect gap is an intrinsic property of the crystal electronic structure rather than of individual configurations. NQEs and explicit BZ sampling therefore act cooperatively but arise from distinct physical mechanisms: the former reshapes the statistical population of proton-sharing configurations, whereas the latter modifies the underlying condensed-phase excitation energies.

To directly assess how thermal disorder, NQEs, and explicit BZ sampling collectively influence agreement with experiment, we compare the ensemble-averaged NH$_4^+$ spectra with the measured absorption spectrum in Fig.~\ref{fig:exp_overlay}. Experimentally, L-pyro-amm exhibits a near-UV absorption onset in the approximate range of $3.8$--$4.5$~eV \cite{stephens2021short}. The classical $\Gamma$-point ensemble substantially overestimates this onset and exhibits strongly suppressed low-energy spectral weight below approximately $5$~eV. Inclusion of NQEs redshifts the ensemble spectrum by approximately $0.64$~eV while simultaneously increasing low-energy optical intensity, demonstrating that quantum proton fluctuations significantly redistribute the statistical population of optically bright configurations.

Together, these results demonstrate that neither thermal sampling alone nor quantum nuclear sampling alone is sufficient to recover the experimental near-UV absorption onset. Accurate agreement with experiment emerges only when both NQEs and explicit BZ effects are incorporated, highlighting the combined importance of quantum nuclear fluctuations and electronic band dispersion for predictive optical spectroscopy of strongly hydrogen-bonded molecular crystals.

\begin{table*}[t]
    \centering
    \caption{\textbf{NQEs increase the population of low-energy bright configurations, while explicit BZ sampling provides an additional redshift of the absorption onset.}
    Summary of the quantum-induced reorganization of the optical excitation landscape and the contribution of explicit Brillouin-zone (BZ) sampling to the absorption onset. $\Delta E_{\mathrm{NQE}}$ denotes the ensemble-averaged redshift of the lowest bright excitation energy upon going from classical to quantum nuclear sampling at the $\Gamma$ point. $\Delta E_{k}$ reports the mean and range of BZ-induced onset redshifts obtained from explicit $k$-point TDDFT benchmarks on representative quantum configurations. Percentages report the fraction of sampled configurations exhibiting at least one bright excitation below 6.0~eV.
}

    \label{tab:summary_optical_reorganization}
    \small
\begin{tabular}{lcccc}
\toprule
Ion &
$\Delta E_{\mathrm{NQE}}$ (eV) &
$\Delta E_{k}$ quantum benchmark range (eV) &
\multicolumn{2}{c}{Bright configurations below 6.0 eV (\%)} \\
 & & & Classical & Quantum \\
\midrule
H$_3$O$^+$ &
$-0.764 \pm 0.031$ &
$-0.85$ [$-0.88$, $-0.82$] &
$2.75 \pm 0.82$ &
$37.50 \pm 2.03$ \\

NH$_4^+$ &
$-0.639 \pm 0.025$ &
$-0.94$ [$-1.00$, $-0.88$] &
$3.00 \pm 0.86$ &
$30.00 \pm 1.82$ \\

K$^+$ &
$-0.530 \pm 0.027$ &
$-1.10$ [$-1.14$, $-1.06$] &
$4.50 \pm 1.03$ &
$23.17 \pm 1.73$ \\

\midrule
All systems &
$-0.644 \pm 0.016$ &
$-0.96$ [$-1.14$, $-0.82$] &
$3.42 \pm 0.52$ &
$30.22 \pm 1.14$ \\

\bottomrule
\end{tabular}
\end{table*}

\section{Discussion}

Low-energy optical absorption in non-aromatic hydrogen-bonded materials is increasingly observed experimentally, yet its microscopic origin and quantitative theoretical description remain unresolved. A central challenge is that the relevant excitation landscape depends sensitively on proton dynamics, hydrogen-bond fluctuations, thermal disorder, and condensed-phase electronic structure. By combining MLIP-enabled long-timescale classical and quantum nuclear sampling with periodic TDDFT calculations, we show that both NQEs and Brillouin-zone (BZ) sampling exert large and comparable influences on the optical absorption onset of a short-hydrogen-bonded molecular crystal.

The central result of this work is that predictive spectroscopy of short-hydrogen-bonded molecular crystals requires extensive quantum nuclear sampling and explicit validation of BZ convergence. Relative to classical ensembles, NQEs stabilize proton-sharing configurations that are strongly suppressed when nuclei are treated classically. This structural reorganization is accompanied by a substantial redistribution of the optical excitation landscape: the lowest bright excitations are redshifted by approximately $0.5$--$0.8~\mathrm{eV}$, while the fraction of configurations exhibiting bright excitations below $6~\mathrm{eV}$ increases from approximately $3\%$ to $30\%$. Thus, the primary role of NQEs is not simply to shift excitation energies, but to reshape the statistical population of optically bright configurations that collectively determine the absorption onset.

The controlled ion substitutions provide further insight into the microscopic origin of these effects. Although K$^+$, NH$_4^+$, and H$_3$O$^+$ generate markedly different local hydrogen-bond environments, all three systems exhibit the same qualitative quantum-induced enhancement of low-energy optical response. The magnitude of the effect follows the degree of local hydrogen-bond cooperativity surrounding the preserved SHB scaffold, with the strongest enhancement observed for H$_3$O$^+$ and the weakest for K$^+$. At the same time, the proton-transfer free-energy landscapes become substantially less ion dependent under quantum nuclear sampling, indicating that proton sharing emerges as a robust feature of the SHB motif itself.

A second major finding is that explicit BZ sampling introduces an additional effect of comparable magnitude to the NQE-induced spectral shift. Despite the large unit cell and molecular character of the crystal, residual band dispersion lowers the optical absorption onset by approximately $0.5$--$1.1~\mathrm{eV}$ in the explicitly BZ-sampled TDDFT benchmark calculations. Importantly, this BZ-induced redshift remains relatively consistent across ion substitutions, ensemble types, and representative configurations, indicating that it reflects the underlying crystal electronic structure rather than a peculiarity of individual nuclear configurations. NQEs and BZ sampling therefore act cooperatively but arise from fundamentally different physical mechanisms: the former redistributes the statistical population of proton-sharing configurations, whereas the latter modifies the reciprocal-space sampling of condensed-phase excitation energies.

Taken together, these effects bring the calculated absorption onset into close agreement with the experimentally observed near-UV range. More broadly, the present results suggest that systematic overestimation of absorption onsets in hydrogen-bonded molecular materials may arise from two approximations that are often made simultaneously: neglect of quantum nuclear fluctuations and insufficient validation of BZ convergence. While the importance of NQEs for hydrogen-bonded systems is well established, the present work demonstrates that ensemble-level optical properties can depend as strongly on configurational statistics and reciprocal-space convergence as on the underlying excited-state methodology itself.

The present study focuses on ground-state ensemble effects and optical absorption. Fluorescence additionally depends on excited-state relaxation pathways, vibronic coupling, nonadiabatic dynamics, and the competition between radiative and nonradiative decay channels, which remain beyond the scope of the present work. Nevertheless, the combination of MLIP-enabled nanosecond-scale classical sampling, hundreds of picoseconds of PIMD sampling, and ensemble excited-state calculations provides a practical route toward predictive spectroscopy in fluctuating condensed-phase systems. The broader question of how widespread comparable NQE- and BZ-induced corrections are across SHB-containing molecular crystals, peptide assemblies, and related biomolecular materials remains an important direction for future work. Because extended hydrogen-bond networks, local electrostatic fields, and, in selected cases, short or low-barrier hydrogen bonds are common motifs in non-aromatic luminophores, future studies may establish whether quantum proton fluctuations and BZ convergence represent general design principles for hydrogen-bond-driven optical response.

\section{Methods}

A multiscale simulation framework combining AIMD, MLIPs, long-timescale classical and quantum nuclear sampling, and ensemble excited-state calculations was employed. The computational protocol consisted of four stages: (i) generation of AIMD reference trajectories for experimentally resolved and ion-substituted crystal structures, (ii) fine-tuning of foundation-model MLIPs to near first-principles accuracy, (iii) long-timescale classical and path-integral molecular dynamics simulations, and (iv) ensemble optical calculations using periodic TDDFT together with explicit BZ benchmark calculations.

\subsection{Crystal structures}

All simulations were based on the experimentally resolved crystal structure of L-pyro-amm (CCDC deposition no.~1981551) \cite{stephens2021short}. The orthorhombic unit cell contains 144 atoms comprising pyroglutamic acid, pyroglutamate, and ammonium ions arranged within a hydrogen-bonded molecular crystal. In addition to conventional N--H--O hydrogen bonds, the structure contains an SHB connecting two carboxylate oxygen atoms with an O--O distance of approximately $2.45$~\AA.

To generate the ion-substituted crystal series, the crystallographic NH$_4^+$ ion was replaced \textit{in silico} by K$^+$ and H$_3$O$^+$ while preserving overall charge neutrality. Atomic positions and lattice vectors were subsequently optimized using density-functional theory without imposing constraints on the hydrogen-bond network or crystal symmetry. The resulting NH$_4^+$, K$^+$, and H$_3$O$^+$ crystal structures were used as starting points for all subsequent molecular dynamics and optical calculations.

Optimized lattice parameters and additional structural benchmarks are provided in Supplementary Note~1.

\subsection{Ab initio molecular dynamics}

AIMD simulations were performed within the Born--Oppenheimer approximation using the \textsc{Quickstep} module of CP2K version 2023.1 \cite{vandevondele2005quickstep}. Electronic states were represented using the Gaussian and plane-wave (GPW) formalism with a DZVP-MOLOPT-SR-GTH basis set together with Goedecker--Teter--Hutter pseudopotentials \cite{goedecker1996separable,vandevondele2007gaussian}. The auxiliary plane-wave density cutoff was set to 320~Ry. Exchange--correlation effects were treated using the BLYP functional together with Grimme D3 dispersion corrections \cite{blyp1,blyp2,grimme2010consistent}. This electronic-structure setup was previously benchmarked for proton-transfer energetics and hydrogen-bond structure in the present SHB system \cite{qaisrani2025acid}.

All simulations employed the DFT-relaxed lattice parameters of the corresponding ion-substituted structures. Trajectories were propagated in the canonical ensemble at 300~K using the velocity-rescaling thermostat with a time constant of 100~fs and a time step of 0.5~fs \cite{bussi2007canonical}. Following equilibration, production trajectories of approximately 50~ps were generated for each composition. The resulting AIMD trajectories served as reference data for subsequent MLIP training and validation.

\subsection{Machine-learned interatomic potentials}

Direct inclusion of NQEs through fully \textit{ab initio} PIMD is computationally prohibitive for the present 144-atom periodic molecular crystals because each nucleus must be represented by multiple path-integral replicas over extended simulation timescales. At the same time, statistically converged optical spectroscopy requires extensive configurational sampling well beyond the timescales accessible to direct AIMD simulations. MLIPs therefore provide a practical route toward long-timescale classical and quantum nuclear sampling while retaining near first-principles accuracy.\cite{behler2007generalized,batatia2022mace}

Composition-specific MLIPs were constructed using the MACE framework (version 0.3.10).\cite{batatia2022mace} The models were initialized from the pretrained MACE-MP-0 foundation model and subsequently fine-tuned using the \texttt{amaceing\_toolkit}.\cite{hanseroth2025amaceingtoolkit} Separate MLIP models were trained for the NH$_4^+$, K$^+$, and H$_3$O$^+$ crystals. For each composition, 400 configurations were extracted from equilibrated AIMD trajectories at intervals of 20~fs in order to sample proton-transfer events, hydrogen-bond fluctuations, and thermally accessible distortions of the SHB network. Fine-tuning employed a combined force--energy loss function with force and energy weights of 10 and 0.1, respectively. The datasets were randomly divided into training and validation subsets, with $10\%$ reserved for validation. Similar MACE-based workflows have previously demonstrated high accuracy for hydrogen-bonded and condensed-phase systems.\cite{grunert2025modeling,hanseroth2025optimizing,flototto2026large,hanseroth2026htscreening,qaisrani2026bridging}

The resulting MLIPs achieved validation energy RMSEs of 1.1--3.7 meV atom$^{-1}$ and force RMSEs of 22--28 meV~\AA$^{-1}$ across the three ion-substituted crystals. Validation against AIMD proton-transfer free-energy profiles together with N--H and O--H radial distribution functions confirmed that the MLIPs accurately reproduce the first-principles proton-transfer landscape and local hydrogen-bond structure governing SHB dynamics. The agreement between MLIP and AIMD free-energy profiles is particularly important because the proton-transfer coordinate serves as the primary descriptor used throughout the present work to characterize proton sharing within the SHB. These validated MLIPs were subsequently employed to generate the long-timescale MLIP MD and MLIP PIMD ensembles used throughout the optical calculations. Detailed training protocols, hyperparameters, RMSE statistics, and structural validation benchmarks are provided in Supplementary Note~2.

\subsection{Classical and path-integral molecular dynamics}

MLIP MD and MLIP PIMD simulations were performed under identical thermodynamic conditions at 300~K with a time step of 0.5~fs.

NQEs were incorporated using the ring-polymer formulation of PIMD \cite{marx1996ab,tuckerman2002path}. Quantum trajectories were propagated using the Path Integral Generalized Langevin Equation Thermostat (PIGLET)  \cite{ceriotti2012efficient}, which accelerates convergence of quantum structural observables with respect to bead number by targeting a colored-noise spectrum matched to the quantum canonical distribution. For hydrogen-bonded systems, PIGLET has been shown to reproduce structural distributions equivalent to those obtained with significantly larger bead numbers in standard PIMD \cite{ceriotti2012efficient}, making it particularly suitable for converging proton-transfer statistics at manageable computational cost. Simulations were implemented using the \textsc{i-PI} package \cite{ceriotti2014pi} interfaced with \textsc{LAMMPS} \cite{LAMMPS}.

Production MLIP MD trajectories of approximately 1~ns and MLIP PIMD trajectories of approximately 300~ps were generated for each ion-substituted crystal. Quantum simulations employed $P = 6$ beads per atom. Additional convergence calculations up to $P = 24$ yielded overlapping proton-transfer free-energy profiles (Supplementary Note~3), confirming structural convergence of the quantum nuclear distributions under the employed PIGLET scheme.

Full bead-number convergence of the ensemble optical spectra was not evaluated directly, as this would require approximately 600 additional hybrid-functional TDDFT calculations at $P = 6$ and several hundred further calculations at higher bead numbers, rendering exhaustive optical convergence testing computationally prohibitive. Instead, the structural proton-transfer distributions, which govern the population of optically bright configurations, were used as the primary convergence diagnostic. The agreement between independent-particle approximation (IPA) and TDDFT ensemble spectra (Supplementary Note~4), which shows that the quantum-induced spectral redistribution primarily reflects nuclear-configuration statistics rather than the specific optical formalism, further supports the use of structural convergence as the appropriate proxy for optical convergence in the present system.

Proton transfer within the SHB was characterized using the coordinate

\begin{equation}
\delta r = d_{\mathrm{O_1-H}} - d_{\mathrm{H-O_2}}
\end{equation}

where $O_1$ and $O_2$ denote the oxygen atoms participating in the SHB. Corresponding free-energy profiles were obtained from the probability distributions according to

\begin{equation}
F(\delta r) = -k_{\mathrm{B}}T \ln P(\delta r) + C.
\end{equation}

Statistical convergence of proton-transfer observables was assessed using block-averaging analysis. The proton-transfer coordinate exhibited a characteristic autocorrelation time of approximately 60~fs for both classical and quantum ensembles. Statistical uncertainties were therefore estimated using a conservative block length of 5~ps, substantially exceeding the proton-transfer decorrelation time. Additional convergence analyses and ensemble statistical benchmarks are provided in Supplementary Notes~3 and~6.

\subsection{Optical calculations}

Optical absorption spectra were computed using the Vienna \textit{Ab initio} Simulation Package (\textsc{VASP}, version 6.5) \cite{Kresse1996,Kresse1999}. Electronic-structure calculations employed the projector augmented-wave (PAW) formalism \cite{Bloechl1994} together with GW PAW pseudopotentials and a plane-wave cutoff energy of 500~eV. Ground-state wave functions and eigenvalues were obtained using the HSE06 hybrid exchange--correlation functional \cite{Heyd2003,Heyd2006}.

For each ion-substituted crystal, optical observables were evaluated independently for every sampled nuclear geometry prior to ensemble averaging. The calculations comprised 400 configurations sampled from the MLIP MD ensembles and 600 bead geometries sampled from the MLIP PIMD ensembles, yielding a total of 1000 optical calculations per composition. The quantum ensemble consisted of 100 independent PIMD frames represented by 6 beads each. Frequency-dependent dielectric functions were evaluated between 0 and 10~eV using 1001 frequency points. Optical spectra were computed using both the independent-particle approximation (IPA) and linear-response TDDFT. IPA calculations were employed as a computationally efficient approach for rapid ensemble-level screening, whereas TDDFT calculations were used for the quantitative excited-state analysis presented in the main text. All calculations employed 480 electronic bands, consisting of 216 occupied and 264 unoccupied states.

The rotationally invariant optical response was obtained from the dielectric tensor according to

\begin{equation}
\overline{\varepsilon}_{\mathrm{inter}}(\omega)
=
\frac{1}{3}
\mathrm{Tr}
\left[
\varepsilon_{\mathrm{inter},\alpha\beta}(\omega)
\right]
\end{equation}

TDDFT calculations were performed by solving the Casida equation starting from the converged HSE06 wave functions and eigenvalues. The Hartree kernel contribution was included using \texttt{LHARTREE = .TRUE.}, and the local exchange--correlation kernel contribution was included using \texttt{LFXC = .TRUE.}. Long-range exchange--correlation kernel and ladder-diagram contributions were not included (\texttt{LADDER = .FALSE.}).

To characterize the ensemble optical response, several configuration-resolved descriptors were evaluated. The primary quantity used throughout the analysis was the lowest bright excitation energy,

\begin{equation}
E_{\mathrm{min}}^{\mathrm{bright}}
=
\min
\left\{
E_n \; | \; f_n > f_{\mathrm{cut}}
\right\},
\end{equation}

where $E_n$ and $f_n$ denote the excitation energy and oscillator strength of transition $n$, respectively, and $f_{\mathrm{cut}} = 0.01$ defines the threshold for optical brightness. This quantity was used to characterize the low-energy absorption onset while excluding optically dark transitions with negligible spectral intensity.

Redistribution of low-energy optical intensity was quantified through the cumulative oscillator strength below 6.0~eV,

\begin{equation}
W_{\mathrm{low}}(6.0)
=
\sum_{E_n \leq 6.0~\mathrm{eV}}
f_n,
\end{equation}

together with corresponding bright-state-restricted quantities obtained by applying the condition $f_n > f_{\mathrm{cut}}$.

The population of low-energy optically bright configurations was quantified through the fraction of configurations containing at least one bright excitation below 6.0~eV,

\begin{equation}
P_{\mathrm{bright}}(<6.0~\mathrm{eV})
=
\frac{
N_{\mathrm{bright}}(<6.0~\mathrm{eV})
}
{N_{\mathrm{tot}}},
\end{equation}

where $N_{\mathrm{tot}}$ is the total number of sampled configurations and $N_{\mathrm{bright}}(<6.0~\mathrm{eV})$ denotes the number of configurations containing at least one bright excitation below 6.0~eV.

For the MLIP PIMD ensembles, optical observables were first evaluated for individual bead geometries and subsequently averaged over the 6 beads associated with each physical PIMD frame before constructing ensemble statistics.

Because statistically converged ensemble hybrid-functional TDDFT calculations with explicit BZ sampling remain computationally prohibitive for the present 144-atom molecular crystals, production optical calculations were performed at the $\Gamma$ point. Additional BZ benchmark calculations were carried out for twelve representative configurations spanning all ion substitutions, ensemble types, and both mean-like and low-onset structures. Selected benchmark configurations were additionally recomputed using the B3LYP hybrid functional \cite{becke1993new} to assess the sensitivity of the optical response to the exchange--correlation functional. Details of the IPA, BZ-sampling, and exchange--correlation functional benchmarks are provided in Supplementary Notes~4 and~5.

\section*{Acknowledgments}

We thank the staff of the Compute Center of the Technische Universit\"at Ilmenau, especially Mr.~Henning Schwanbeck, for providing an excellent research environment.
This work was supported by the Carl-Zeiss-Stiftung through SustEnMat (funding code P2023-02-008), the Th\"uringer Aufbaubank (TAB) through KapMemLyse (grant no.~2024 FGR 0081/0082), and the European Social Fund Plus (ESF+).

\section*{Competing interests}

The authors declare no competing interests.

\section*{Code availability}

Simulated structures, representative trajectories, processed data, and analysis scripts will be made available in a public repository. The simulations used publicly available codes, including CP2K, i-PI, LAMMPS, and VASP, as described in the Methods. The MLIP fine-tuning protocol was adapted from Ref.~\cite{hanseroth2025amaceingtoolkit}.

\section*{Author contributions}

M.N.Q.~conceived the idea.
J.H.~and M.N.Q.~performed the AIMD calculations.
J.H.~trained the MLIP models and performed the ML MD and ML PIMD simulations.
M.~Gro{\ss}mann performed the IPA and TDDFT excited-state calculations.
M.N.Q., J.H., M.~Grunert, and M.~Gro{\ss}mann analyzed and visualized the results.
M.N.Q.~wrote the first draft of the manuscript.
E.R., C.D., and M.N.Q.~supervised the work.
All authors discussed the results, revised the manuscript, and approved the final version.

\bibliography{literature}

\clearpage
\onecolumngrid
\clearpage

\pagestyle{fancy}
\fancyhf{}
\fancyhead[L]{\textbf{Supporting Information}}
\renewcommand{\headrulewidth}{0.4pt}
\markboth{Supporting Information}{Supporting Information}

\begin{center}
\vspace*{1.2cm}

{\Large \textbf{Supporting Information}}\\[0.6cm]

{\large \textbf{Quantum nuclear and band-dispersion effects recover near-UV absorption in short-hydrogen-bonded organic crystals}}\\[1.0cm]

\large
Jonas H{\"a}nseroth,$^{1}$ Max Gro{\ss}mann,$^{1}$ Malte Grunert,$^{1}$\\
Ali Hassanali,$^{2}$ Erich Runge,$^{1}$ Christian Dre{\ss}ler,$^{1}$\\
and Muhammad Nawaz Qaisrani$^{1,*}$\\[0.7cm]

\normalsize
$^{1}$Institute of Physics and Institute of Micro- and Nanotechnologies,\\
Technische Universit\"at Ilmenau, 98693 Ilmenau, Germany\\[0.3cm]

$^{2}$Condensed Matter and Statistical Physics,\\
The Abdus Salam International Centre for Theoretical Physics (ICTP),\\
34151 Trieste, Italy\\[0.6cm]

$^{*}$Corresponding author: muhammad-nawaz.qaisrani@tu-ilmenau.de\\[1.0cm]

\hrule

\end{center}

\vspace{0.8cm}

\setcounter{section}{0}
\renewcommand{\thesection}{S\arabic{section}}

\setcounter{figure}{0}
\renewcommand{\thefigure}{S\arabic{figure}}

\setcounter{table}{0}
\renewcommand{\thetable}{S\arabic{table}}

\setcounter{equation}{0}
\renewcommand{\theequation}{S\arabic{equation}}

\section*{Supplementary Note 1: Lattice response to ion substitution}

To probe the structural adaptability of the short-hydrogen-bond (SHB) framework, the native ammonium (NH$_4^+$) cation was replaced computationally by potassium (K$^+$) and hydronium (H$_3$O$^+$) ions, followed by unconstrained variable-cell density-functional-theory (DFT) optimizations. All structural optimizations were performed using the same CP2K setup employed for the \textit{ab initio} molecular dynamics simulations described in the Methods section, including the exchange--correlation functional, basis sets, pseudopotentials, dispersion correction, and numerical parameters.

In all cases, the SHB motif was preserved and the orthorhombic crystal symmetry was retained without inducing large-scale structural reconstruction.

The DFT-optimized structure of the native NH$_4^+$ crystal, obtained at 0~K, exhibits a contraction relative to the room-temperature experimental lattice parameters. The optimized unit-cell volume is $1056.6~\text{\AA}^3$, compared with the experimental value of $1277.4~\text{\AA}^3$, with the largest deviation occurring along the crystallographic \emph{a} axis. This difference reflects the comparison between a static 0~K DFT reference and a finite-temperature experimental structure and may additionally depend on the chosen exchange--correlation functional.

Against this theoretical reference, the K$^+$ and H$_3$O$^+$ substituted structures display closely similar lattice metrics. Their optimized volumes, $1049.9~\text{\AA}^3$ for K$^+$ and $1036.9~\text{\AA}^3$ for H$_3$O$^+$, differ by only approximately $2\%$ from the optimized NH$_4^+$ structure. These results indicate that cation substitution induces only minor changes in global lattice parameters, with structural accommodation occurring primarily through local rearrangements of coordination and hydrogen-bond geometry.

To assess finite-temperature stability and the robustness of the optimized lattice parameters, additional \textit{ab initio} NPT molecular dynamics simulations were performed for both the experimental and DFT-optimized NH$_4^+$ crystal structures using the same CP2K protocol employed throughout this work. Pressure control was carried out using the MTTK barostat.\cite{martyna1996explicit} As shown in Supplementary Fig.~\ref{fig:aimd_npt_volume_evolution}, the trajectory initiated from the experimental unit cell undergoes a gradual volume contraction, whereas the trajectory initiated from the optimized structure remains close to its initial volume. In both cases, the cell volumes converge toward a similar equilibrium value without disruption of the SHB framework. These calculations indicate that the DFT-optimized structure corresponds closely to the finite-temperature equilibrium volume predicted by the employed electronic-structure method.

The consistent contraction relative to experiment therefore reflects a common theoretical reference state rather than ion-specific structural distortions. Because the present ion-substitution series is designed to compare internally consistent nuclear and optical responses rather than to reproduce absolute thermal expansion, all compositions were analysed within the same DFT-relaxed reference protocol. The optimized NH$_4^+$, K$^+$, and H$_3$O$^+$ crystal structures consequently provide an internally consistent basis for investigating the influence of ion substitution on proton-transfer dynamics, nuclear quantum effects, and optical response.

\begin{table}[ht]
\centering
\caption{
\textbf{Ion substitution induces only minor lattice distortions.}
First-principles optimized lattice parameters of the ion-substituted crystals. All optimized cells retain orthorhombic symmetry ($\alpha=\beta=\gamma=90^\circ$). Experimental parameters are listed for reference.
}
\begin{tabular}{lcccc}
\toprule
Ion type & $a$ ({\AA}) & $b$ ({\AA}) & $c$ ({\AA}) & Volume ({\AA}$^3$) \\
\midrule
NH$_4^+$   & 4.3152 & 14.0831 & 17.3868 & 1056.6283 \\
K$^+$      & 4.3336 & 14.3136 & 16.9253 & 1049.8596 \\
H$_3$O$^+$ & 4.2927 & 13.6604 & 17.6829 & 1036.9139 \\
\midrule
Experiment & 5.1513 & 14.5596 & 17.0315 & 1277.3773 \\
\bottomrule
\end{tabular}
\label{tab:cellopt}
\end{table}

\begin{figure*}[ht]
\centering
\includegraphics{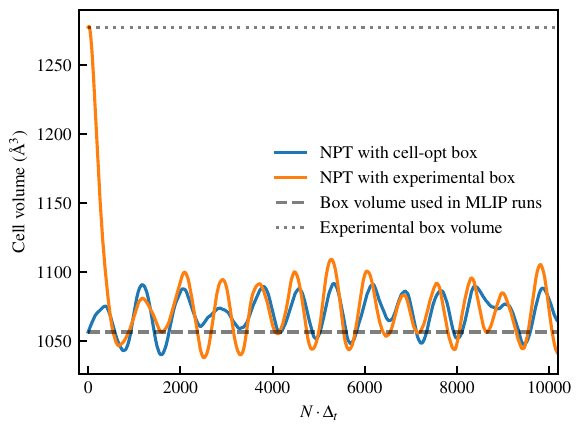}
\caption{\textbf{The optimized crystal volume remains stable under finite-temperature NPT dynamics.}
Evolution of the cell volume during finite-temperature \textit{ab initio} NPT molecular dynamics simulations starting from the experimental and cell-optimized NH$_4^+$ simulation boxes. The trajectory initiated from the experimental cell contracts toward the same theoretical equilibrium volume, whereas the trajectory initiated from the optimized structure remains close to its initial volume. In both cases, the SHB framework remains intact.
}
\label{fig:aimd_npt_volume_evolution}
\end{figure*}

\clearpage

\section*{Supplementary Note 2: MLIP fine-tuning and validation}

Composition-specific MLIPs were constructed using the MACE framework (version 0.3.10).\cite{batatia2022mace} The pretrained MACE-MP-0 foundation model was fine-tuned using the \texttt{amaceing\_toolkit} workflow.\cite{hanseroth2025amaceingtoolkit} Separate MLIP models were trained for the NH$_4^+$, K$^+$, and H$_3$O$^+$ crystals.

For each composition, 400 configurations were extracted from equilibrated AIMD trajectories at intervals of $20~\mathrm{fs}$ to sample proton-transfer events, hydrogen-bond fluctuations, and thermally accessible distortions of the SHB network. Fine-tuning employed a combined force--energy loss function with force and energy weights of 10 and 0.1, respectively. The datasets were randomly divided into training and validation subsets, with $10\%$ reserved for validation. Supplementary Table~\ref{tab:mace_ft} summarizes the training protocol and hyperparameters used during MLIP fine-tuning.

\begin{table}[h]
\centering
\caption{\textbf{Fine-tuning parameters used for MACE-MP-0 optimization.}
Training protocol and hyperparameters used for fine-tuning the pretrained MACE-MP-0 foundation model on AIMD reference data for the ion-substituted crystals.}
\label{tab:mace_ft}
\begin{tabular}{lcccccc}
\toprule
Foundation model & Force weight & Energy weight & Training set & Validation fraction & Batch size & Epochs \\
\midrule
MACE-MP-0 & 10 & 0.1 & 400 frames, $\Delta t=20~\mathrm{fs}$ & 0.1 & 5 & 200 \\
\bottomrule
\end{tabular}
\end{table}

The resulting models achieved low energy and force errors across both training and validation datasets, indicating stable interpolation of the AIMD reference data without significant overfitting. Validation energy RMSEs ranged from $1.1$ to $3.7~\mathrm{meV~atom^{-1}}$, while validation force RMSEs ranged from $22$ to $28~\mathrm{meV}\,\text{\AA}^{-1}$ across the three ion-substituted crystals (Supplementary Table~\ref{tab:mace_error}).

\begin{table}[h]
\centering
\caption{\textbf{MLIPs reproduce AIMD energies and forces across all ion substitutions.}
Training and validation root-mean-square errors (RMSEs) of energies and forces for the fine-tuned MLIPs of the NH$_4^+$, K$^+$, and H$_3$O$^+$ crystals.}
\label{tab:mace_error}
\begin{tabular}{lccc}
\toprule
Quantity & NH$_4^+$ & K$^+$ & H$_3$O$^+$ \\
\midrule
\multicolumn{4}{l}{\textit{Training set}} \\
RMSE $E$ (meV atom$^{-1}$) & 2.2 & 1.1 & 3.8 \\
RMSE $F$ (meV \AA$^{-1}$) & 20.9 & 20.3 & 25.4 \\
\midrule
\multicolumn{4}{l}{\textit{Validation set}} \\
RMSE $E$ (meV atom$^{-1}$) & 2.2 & 1.1 & 3.7 \\
RMSE $F$ (meV \AA$^{-1}$) & 22.0 & 22.6 & 28.0 \\
\bottomrule
\end{tabular}
\end{table}

\begin{figure*}[ht]
\centering
\includegraphics{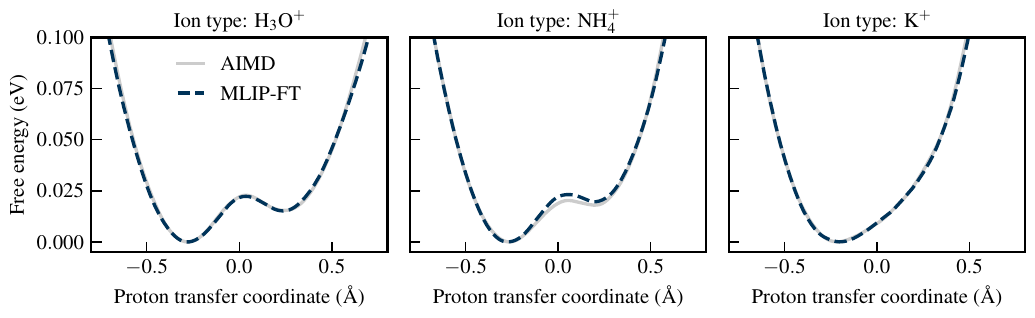}
\caption{\textbf{MLIPs reproduce AIMD proton-transfer free-energy profiles.}
Comparison of proton-transfer free-energy profiles obtained from AIMD and MLIP MD simulations for L-pyro-H$_3$O$^+$, L-pyro-NH$_4^+$, and L-pyro-K$^+$. The close agreement demonstrates that the MLIPs reproduce the first-principles proton-sharing free-energy landscape governing SHB fluctuations.
}
\label{fig:mlip_fe}
\end{figure*}

\begin{figure*}[ht]
\centering
\includegraphics{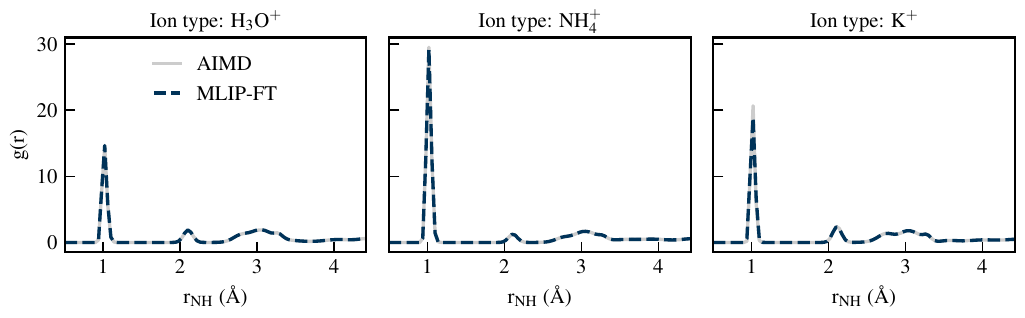}
\caption{\textbf{MLIPs reproduce local N--H hydrogen-bond structure.}
Comparison of N--H radial distribution functions obtained from AIMD and MLIP MD simulations for the ion-substituted crystals. The agreement confirms that the fine-tuned MLIPs reproduce the local hydrogen-bond structure and proton environments sampled in the first-principles trajectories.
}
\label{fig:mlip_nh}
\end{figure*}

\begin{figure*}[ht]
\centering
\includegraphics{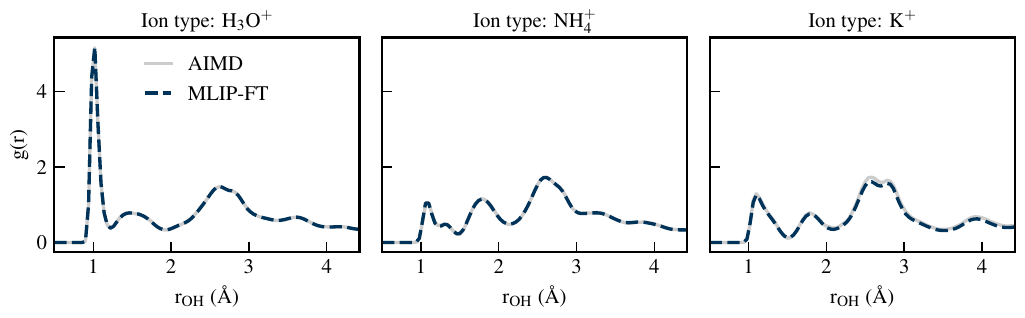}
\caption{\textbf{MLIPs reproduce SHB O--H distributions.}
Comparison of O--H radial distribution functions obtained from AIMD and MLIP MD simulations for the ion-substituted crystals. The MLIP trajectories reproduce the structural fluctuations and proton-sharing behaviour associated with the SHB motif.
}
\label{fig:mlip_oh}
\end{figure*}

The accuracy of the MLIPs was further assessed using structural observables directly relevant to SHB proton dynamics. Validation against AIMD reference trajectories was performed using proton-transfer free-energy profiles together with N--H and O--H radial distribution functions. Supplementary Figs.~\ref{fig:mlip_fe}--\ref{fig:mlip_oh} show close agreement between MLIP and AIMD structural observables across all three ion substitutions, confirming that the MLIPs reproduce both the proton-transfer landscape and the local hydrogen-bond structure of the underlying first-principles simulations.

\clearpage

\section*{Supplementary Note 3: PIMD bead-number convergence}

To assess convergence of the quantum structural distributions, the number of path-integral beads was varied from $P=6$ to $P=24$. For each bead number, MLIP-based PIMD simulations were performed and proton-transfer free-energy profiles were computed for the SHB coordinate.

As shown in Supplementary Fig.~\ref{fig:bead_convergence}, the free-energy profile obtained using $P=6$ closely overlaps with the profiles obtained using larger bead numbers, including $P=24$. These results confirm convergence of the proton-transfer distributions under the employed PIGLET scheme and support the use of $P=6$ for the production quantum simulations reported in the main text.

\begin{figure*}[ht]
\centering
\includegraphics{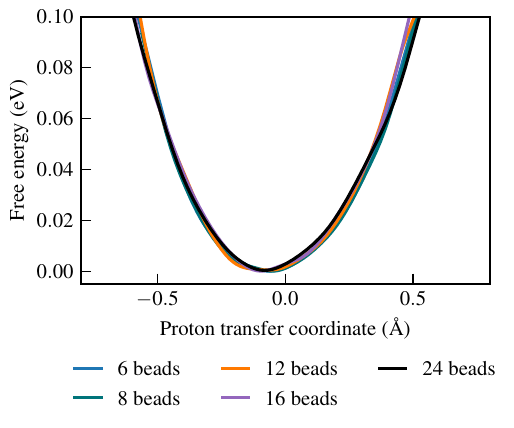}
\caption{\textbf{Six PIGLET beads are sufficient to converge the proton-transfer free-energy profile.}
Convergence of proton-transfer free-energy profiles with respect to the number of PIMD beads for L-pyro-NH$_4^+$. The free-energy landscapes obtained using $P=6$ and larger bead numbers exhibit close overlap, indicating convergence of the quantum structural distributions under the PIGLET thermostat scheme at $300~\mathrm{K}$.
}
\label{fig:bead_convergence}
\end{figure*}

\clearpage

\section*{Supplementary Note 4: IPA and TDDFT ensemble optical calculations}

Independent-particle approximation (IPA) calculations were initially employed as a computationally efficient approach for analysing large classical and quantum nuclear ensembles and for assessing whether nuclear quantum fluctuations qualitatively redistribute low-energy optical intensity. Because IPA calculations are substantially less expensive than hybrid-functional TDDFT, they enabled rapid screening of ensemble-level optical trends prior to the more demanding excited-state calculations.

To determine whether the observed quantum-induced spectral redistribution depends sensitively on the optical-response formalism, the same nuclear ensembles were subsequently analysed using TDDFT. Although IPA neglects explicit electron--hole interactions and is not used as the quantitative excited-state reference, the qualitative enhancement of low-energy optical intensity under quantum nuclear sampling remains unchanged across all ion substitutions.

\begin{figure*}[ht]
\centering
\includegraphics{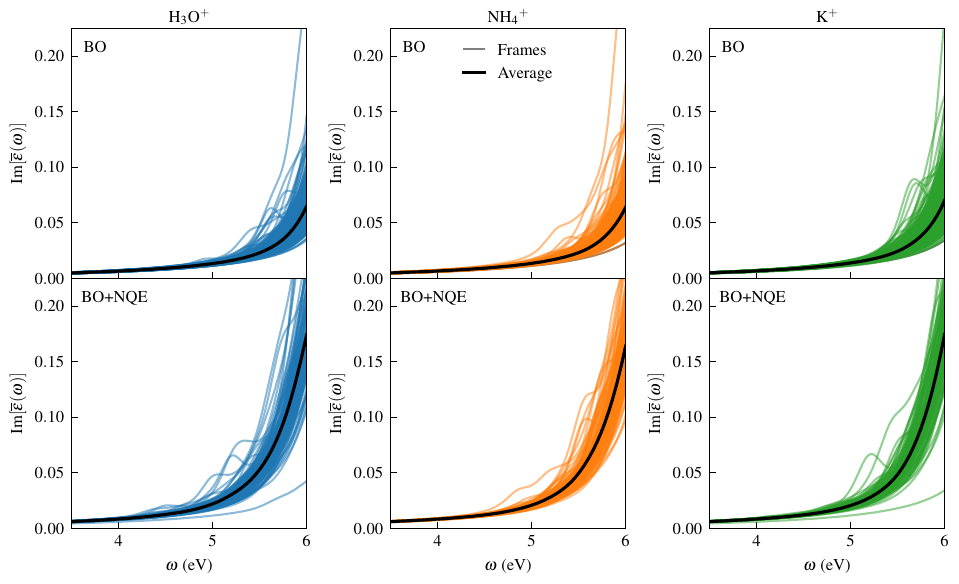}
\caption{\textbf{IPA and TDDFT predict similar quantum-induced enhancement.}
Imaginary part of the orientationally averaged dielectric function computed within the independent-particle approximation for structures sampled from classical MLIP MD (top row) and quantum MLIP PIMD (bottom row) molecular dynamics for $\mathrm{H_3O^+}$, $\mathrm{NH_4^+}$, and $\mathrm{K^+}$ crystals. Thin lines denote spectra of individual configurations and thick black curves represent ensemble averages. Quantum nuclear sampling systematically enhances the statistical population of low-energy optically bright configurations below approximately $5~\mathrm{eV}$ across all ion substitutions.
}
\label{fig:ipa_spectra}
\end{figure*}

Supplementary Fig.~\ref{fig:ipa_spectra} compares the ensemble-resolved IPA spectra obtained from classical and quantum nuclear ensembles. Similar to the TDDFT results discussed in the main text, quantum nuclear sampling systematically increases the statistical population of low-energy optically bright configurations below approximately $5~\mathrm{eV}$. The persistence of the low-energy optical enhancement across both IPA and TDDFT calculations indicates that the dominant origin of the observed spectral redshift is the quantum redistribution of sampled nuclear configurations rather than the specific optical-response formalism.

\clearpage

\section*{Supplementary Note 5: Brillouin-zone sampling and functional dependence of the optical spectra}

\begin{figure*}[ht]
\centering
\includegraphics[width=0.85\textwidth]{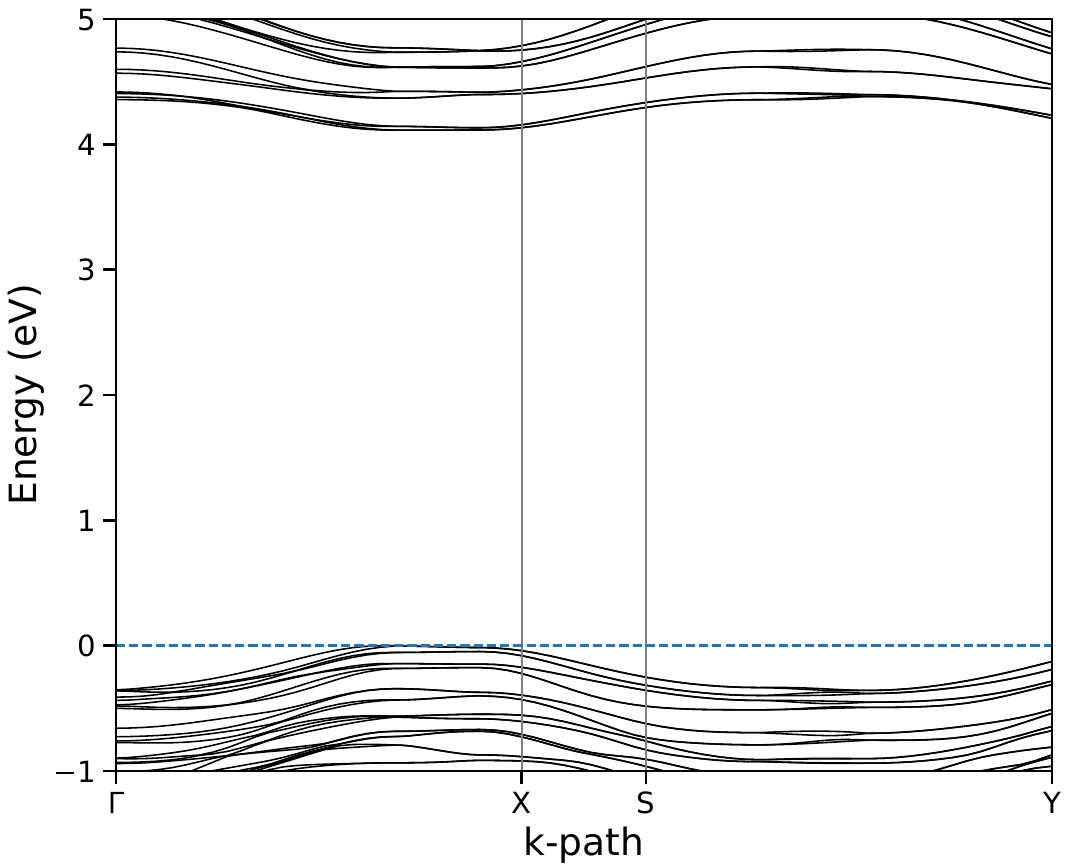}
\caption{\textbf{The electronic gap exhibits weak indirect character at the PBE+D3 level.}
Electronic band structure of the NH$_4^+$ crystal computed at the PBE+D3 level. The band-edge states exhibit weak indirect character, with the valence-band maximum and conduction-band minimum located slightly away from the $\Gamma$ point along the $\Gamma$--X direction. This calculation is used as a qualitative diagnostic of residual reciprocal-space dispersion; quantitative BZ-induced optical shifts are obtained from explicit BZ-sampled HSE06/TDDFT benchmark spectra.
}
\label{fig:band_structure}
\end{figure*}

Because production ensemble calculations were performed at the $\Gamma$ point, additional calculations employing explicit Brillouin-zone (BZ) sampling were carried out to assess the influence of reciprocal-space convergence on the optical spectra. A PBE-level band-structure calculation reveals residual band-edge dispersion and weak indirect electronic character away from the $\Gamma$ point (Supplementary Fig.~\ref{fig:band_structure}), providing the physical motivation for the BZ benchmarks discussed in the main text.

The PBE-level band-structure calculation is used here only as a qualitative diagnostic of residual reciprocal-space dispersion. The quantitative BZ-induced optical shifts discussed in the main text and reported below are obtained directly from explicit $\Gamma$-point and BZ-sampled HSE06/TDDFT spectra computed for representative configurations.

\begin{figure*}[ht]
\centering
\includegraphics{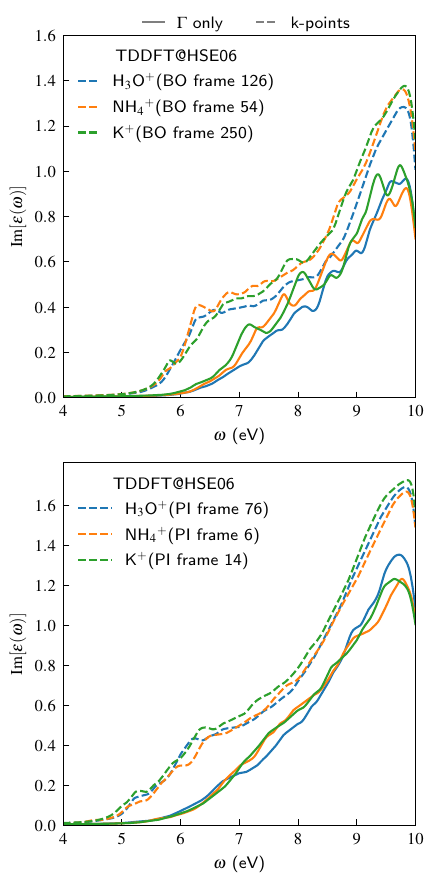}
\caption{\textbf{Explicit Brillouin-zone sampling systematically redshifts the optical onset.}
Comparison of $\Gamma$-point and explicitly BZ-sampled TDDFT absorption spectra for configurations representative of the ensemble-average optical response. Explicit BZ sampling lowers the absorption onset and enhances low-energy spectral weight across all ion substitutions and for both classical (top) and quantum (bottom) nuclear ensembles.
}
\label{fig:kpoint_mean}
\end{figure*}

Representative configurations were selected from both classical and quantum nuclear ensembles for each ion substitution. As discussed in the main text, two classes of structures were analysed: configurations representative of the ensemble-average optical response and configurations exhibiting the lowest-energy absorption onset within the sampled ensembles. Supplementary Fig.~\ref{fig:kpoint_mean} compares the corresponding $\Gamma$-point and explicitly BZ-sampled TDDFT spectra for the ensemble-average structures.

\noindent
To quantify the effect of reciprocal-space sampling on the optical response, selected configurations were recomputed using both $\Gamma$-point and explicit BZ sampling. The quantities reported in Supplementary Table~\ref{tab:si_kpoint_benchmarks} were obtained from direct comparison of the corresponding HSE06/TDDFT spectra.

The BZ-induced absorption-onset shift was defined as
\begin{equation}
\Delta E_{\mathrm{onset}} = E_{\mathrm{onset}}^{k} - E_{\mathrm{onset}}^{\Gamma},
\end{equation}
where $E_{\mathrm{onset}}^{\Gamma}$ and $E_{\mathrm{onset}}^{k}$ denote the absorption onset obtained from the $\Gamma$-point and explicitly BZ-sampled spectra, respectively. Negative values therefore correspond to a redshift upon reciprocal-space sampling.

The enhancement of low-energy spectral weight upon explicit BZ sampling was quantified through
\begin{equation}
R_{\mathrm{weight}} = \frac{W_{\mathrm{low}}^{k}}{W_{\mathrm{low}}^{\Gamma}},
\end{equation}
where
\begin{equation}
W_{\mathrm{low}} = \int_{0}^{E_{\mathrm{cut}}} I(E)\, dE
\end{equation}
denotes the integrated spectral weight below a fixed low-energy cutoff $E_{\mathrm{cut}}$. Here $E_{\mathrm{cut}} = 6~\mathrm{eV}$, defining all energies below $6~\mathrm{eV}$ as the low-energy region. $I(E)$ denotes the absorption intensity.

\begin{table*}[t]
\centering
\caption{
\textbf{Explicit Brillouin-zone sampling consistently redshifts the optical onset.}
Full set of explicit $k$-point HSE06/TDDFT benchmark calculations used to quantify BZ-induced spectral shifts. $\Delta E_{\mathrm{onset}}$ denotes the shift in absorption onset upon going from $\Gamma$-point to explicit BZ sampling. The weight ratio quantifies the corresponding enhancement of low-energy spectral weight, defined as the integrated intensity below $6~\mathrm{eV}$.
}
\label{tab:si_kpoint_benchmarks}
\begin{tabular}{llrr}
\toprule
System & Ensemble / frame & $\Delta E_{\mathrm{onset}}$ (eV) & Weight ratio \\
\midrule
H$_3$O$^+$ & ML MD mean & $-0.75$ & $5.40$ \\
H$_3$O$^+$ & ML MD low-onset & $-0.65$ & $4.35$ \\
H$_3$O$^+$ & ML PIMD mean & $-0.88$ & $5.49$ \\
H$_3$O$^+$ & ML PIMD low-onset & $-0.82$ & $4.36$ \\
K$^+$ & ML MD mean & $-0.64$ & $4.80$ \\
K$^+$ & ML MD low-onset & $-0.56$ & $2.87$ \\
K$^+$ & ML PIMD mean & $-1.06$ & $7.18$ \\
K$^+$ & ML PIMD low-onset & $-1.14$ & $7.45$ \\
NH$_4^+$ & ML MD mean & $-0.70$ & $5.17$ \\
NH$_4^+$ & ML MD low-onset & $-0.46$ & $2.79$ \\
NH$_4^+$ & ML PIMD mean & $-1.00$ & $5.78$ \\
NH$_4^+$ & ML PIMD low-onset & $-0.88$ & $4.87$ \\
\bottomrule
\end{tabular}
\end{table*}

\begin{figure*}[ht]
\centering
\includegraphics{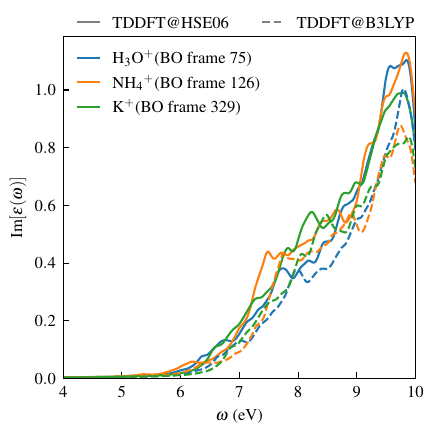}
\caption{
\textbf{A selected B3LYP benchmark preserves the qualitative low-energy optical enhancement.}
Comparison of HSE06 and B3LYP absorption spectra for one representative NH$_4^+$ quantum configuration close to the ensemble-average optical response. B3LYP produces a modest blue shift relative to HSE06 while preserving the qualitative low-energy optical enhancement.
}
\label{fig:func_bench}
\end{figure*}

To assess the sensitivity of the optical spectra to the exchange--correlation functional, one representative NH$_4^+$ quantum configuration close to the ensemble-average optical response was recomputed using the B3LYP hybrid functional.\cite{becke1993new} Supplementary Fig.~\ref{fig:func_bench} compares the resulting HSE06 and B3LYP spectra. B3LYP produces a modest blue shift relative to HSE06 while preserving the enhanced low-energy optical intensity and the qualitative quantum-induced redistribution of spectral weight. This selected benchmark indicates that the qualitative low-energy optical enhancement is not an artefact of the specific hybrid functional used for the main HSE06 calculations.

\clearpage

\section*{Supplementary Note 6: Statistical convergence and suppression of ion-dependent proton localization}

To assess statistical convergence and quantify the suppression of ion-dependent proton-localization asymmetry by nuclear quantum effects, we analysed long-timescale MLIP MD and MLIP PIMD trajectories using block-averaging statistics. The proton-transfer coordinate $\delta r$ exhibited a characteristic autocorrelation time of approximately $60~\mathrm{fs}$ for both classical and quantum ensembles. Statistical uncertainties were therefore estimated using block averaging with a conservative block length of $5~\mathrm{ps}$, substantially exceeding the proton-transfer decorrelation time.

Supplementary Table~\ref{tab:ion_dependence_reduction} summarizes the ion-dependent spread of proton-transfer descriptors across the three ion substitutions. Under classical sampling, the proton-localization magnitude $\langle |\delta r| \rangle$ and proton-transfer fluctuation width $\sigma(\delta r)$ vary substantially between ions. Inclusion of nuclear quantum effects strongly suppresses this variability. The ion-dependent spread decreases by approximately $60\%$ for both $\langle |\delta r| \rangle$ and $\sigma(\delta r)$, indicating that quantum fluctuations substantially suppress ion-dependent proton-localization asymmetry within the SHB motif.

\begin{table*}[ht]
\centering
\caption{
\textbf{Nuclear quantum effects suppress ion-dependent proton localization.}
Reduction of ion-dependent structural variability upon inclusion of nuclear quantum effects. Ion dependence is quantified from the standard deviation of the ion-resolved ensemble averages. The absolute decrease is reported as the positive difference between the classical and quantum ion-dependent spreads.
}
\label{tab:ion_dependence_reduction}
\small
\begin{tabular}{lcccc}
\toprule
Observable & Ion spread (MLIP MD) & Ion spread (MLIP PIMD) & Absolute decrease & (MLIP PIMD)/(MLIP MD) ratio \\
\midrule
$\langle |\delta r| \rangle$ (\AA) & $0.0304$ & $0.0115$ & $0.0189$ & $0.38$ \\
$\sigma(\delta r)$ (\AA) & $0.0368$ & $0.0159$ & $0.0209$ & $0.43$ \\
$\langle R_{\mathrm{OO}} \rangle$ (\AA) & $0.0076$ & $0.0044$ & $0.0033$ & $0.57$ \\
\bottomrule
\end{tabular}
\end{table*}


\end{document}